\documentstyle[aps,epsf,eqsecnum]{revtex}
\begin{document}
\preprint{UAHEP 983}
\draft
\title{Charged Dilatonic Black Holes: String frame vs Einstein frame}
\author{Roberto Casadio}
\address{Dipartimento di Fisica, Universit\`a di Bologna, and \\
I.N.F.N., Sezione di Bologna, \\
via Irnerio 46, 40126 Bologna, Italy}
\author{Benjamin Harms}
\address{Department of Physics and Astronomy,
The University of Alabama\\
Box 870324, Tuscaloosa, AL 35487-0324}
\maketitle
\begin{abstract}
The descriptions of Reissner-Nordstr\"om and Kerr-Newman dilatonic
black holes in the Einstein frame are compared to those in the string
frame.
We describe various physical measurements in the two frames and show
which experiments can distinguish between the two frames.  
In particular we discuss the gyromagnetic ratios of black holes, 
the decay law via Hawking radiation and the propagation of light 
on black hole backgrounds.
\end{abstract}
\section{Introduction}
Superstring theory \cite{gsw}
compactified down to physical four space-time dimensions is generally
accepted as the description of curved backgrounds as non-vanishing 
expectation values of massless string excitations in a language which 
reproduces Einstein's general relativity.
However it is not {\em a priori} clear in which frame such a description
should be better formulated, since there are several frameworks possible.
\par
Specifically, upon compactification of extra dimensions and nearly
independently of both the content of the starting theory and the
compactification scheme, one expects to obtain a low-energy, tree-level,
effective 4-dimensional action $S_\sigma$ for the evolution of the
uncompactified degrees of freedom of the string, which is in the form of a
conformally invariant non-linear $\sigma$-model.
A spin 2 gravitational field $G_{ij}$
(Latin indices run from 0 to 3) and an antisymmetric field $B_{ij}$
which couple respectively to the string vertex operators for their
own emission (see \cite{lovelace,callan} and Refs. therein), emerge in a
natural way in $S_\sigma$
\begin{equation}
S_\sigma={1\over 2\,\lambda_s^2}\,\int d^4x\,\left[h^{\alpha\beta}\,
G_{ij}\,\partial_\alpha x^i\,\partial_\beta x^j
+\epsilon^{\alpha\beta}\,B_{ij}\,\partial_\alpha x^i\,\partial_\beta x^j
+\ldots\right]
\ ,
\label{s}
\end{equation}
where $\lambda_s$ is the string length, $h_{\alpha\beta}$ the world-sheet
metric tensor, $\epsilon^{\alpha\beta}$ the Levi-Civita symbol in two
dimensions and ellipses stand for fermionic terms as well as gauge fields
and other moduli.
\par
Of course were one able to perform the derivation completely, the
background expectation values of the various fields could be fully
determined from the action of the fundamental theory from which one
has started.
This purely descriptive picture is hard to perform in general,
if possible at all.
The most serious obstacle to obtaining such a picture is that there are at
present no models which describe
compactification as a dynamical process taking place in the higher
dimensional space of (super)string theory and leading to an evolution
towards the present state of the Universe as the preferred vacuum
(for a review of early attempts see {\em e.g.} \cite{kolb}).
Thus one is naturally led to seek other ways of solving for the fields.
A possible guide is the requirement of conformal invariance on the
world-sheet
for a generalized $D$-dimensional $\sigma$-model which furnishes us a
complete
set of renormalization equations.
Further, it turns out that in the latter set one has also to include the
scalar excitation $\phi$ (the dilaton) which couples to the world-sheet
scalar curvature \cite{callan}.
\par
The aforementioned equations, supplemented with suitable boundary
conditions, are sufficient to determine the background fields without
making any further reference to string theory.
This is a result of the fact that the same equations can also be obtained
by varying an effective action which, for the particular case of
$B_{ij}\equiv 0$ and non zero Maxwell strength tensor of
electrodynamics $F_{ij}$, can be written (see \cite{horo}, but we work
in four space-time dimensions)
\begin{equation}
S_{SF}={1\over 2}\,\int d^4 x\,\sqrt{-G}\,e^{-\phi}\,\left[
{1\over \lambda_s^2}\,\left(R_{(G)}
+G^{ij}\,\nabla_i\phi\,\nabla_j\phi\right)
-{1\over\alpha^2}\,e^{(1-a)\,\phi}\,F^2\right]
\ ,
\label{s_s}
\end{equation}
where $R_{(G)}$ is the scalar curvature of the metric $G_{ij}$,
$\alpha$ is the electromagnetic coupling constant and
$a$ is the dilaton coupling constant with the dilaton assumed
to remain massless \cite{a}.
This is the effective action which describes the background fields of
the selected string vacuum in the so called {\em string frame} (SF).
The name is justified by the fact that the uncompactified degrees of
freedom of the string move along a geodesic of the metric $G_{ij}$ as
can be inferred from (\ref{s}).
The equations of motion following from (\ref{s_s}) can be written as
(we omit the subscript $(G)$ and set $\lambda_s=\alpha=1$)
\begin{eqnarray}
&&
R_{ij}-{1\over 2}\,G_{ij}\,R+{1\over 2}\,G_{ij}\,\left(\nabla\phi\right)^2
-G_{ij}\,\nabla^2\phi+\nabla_i\nabla_j\phi
-2\,e^{(1-a)\,\phi}\,T_{ij}^{EM}=0
\nonumber \\
&&
\nabla^2\phi-\left(\nabla\phi\right)^2+a\,e^{(1-a)\,\phi}\,F^2=0
\nonumber \\
&&
\nabla_i\left(e^{-a\,\phi}\,F^{ij}\right)=0
\ ,
\label{eq_s}
\end{eqnarray}
where $\nabla$ denotes the covariant derivative with respect to
the metric $G_{ij}$ and the electromagnetic energy-momentum tensor
is given as
\begin{eqnarray}
T^{EM}_{ij} = F_{ik}\,F^k_j-{1\over{4}}\,G_{ij}\,F^2
\ .
\label{t_em}
\end{eqnarray}
\par
The effective action and equations of motion can be further modified
by applying the following conformal transformation
\begin{equation}
G_{ij}=e^{\phi-\phi_0}\,g_{ij}
\ ,
\label{conf}
\end{equation}
where the constant $\phi_0$ can be taken to be the average value of
the dilaton in the present Universe.
Since in these notes we consider only asymptotically flat
cases, we will assume $\phi_0\equiv 0$ in the forthcoming sections.
We further define the Planck length as
\begin{eqnarray}
\ell_p^2=e^{\phi_0}\,\lambda_s^2
\ ,
\end{eqnarray}
thus obtaining the action in the {\em Einstein frame} (EF),
\begin{equation}
S_{EF}={1\over 2}\int d^4x \,\sqrt{-g}\,\left[
{1\over\ell_p^2}\,\left(R_{(g)}
-{1\over 2}\,g^{ij}\,\nabla_i\phi\,\nabla_j\phi\right)
-{1\over\alpha^2}\,e^{-a\,\phi}\,F^2\right]
\ ,
\label{s_e}
\end{equation}
in which we point out that space-time coordinates have been left
unchanged and $R_{(g)}$ is the curvature related to $g_{ij}$.
The new equations of motion are obtained by applying the
same mapping to the previous ones given in (\ref{eq_s})
(again we omit the subscript $(g)$ and set $\ell_p=\alpha=1$),
\begin{eqnarray}
&&
R_{ij} = {1\over{2}}\,\nabla_i\phi\,\nabla_j\phi + 2\,
e^{-a\,\phi}\,T^{EM}_{ij}
\nonumber \\
&&
\nabla^2\phi+a\,e^{-a\,\phi}\,F^2=0
\nonumber \\
&&
\nabla_i(e^{-a\,\phi}\,F^{ij}) = 0
\ ,
\label{eq_e}
\end{eqnarray}
where $\nabla$ is now the covariant derivative with respect
to the metric $g_{ij}$.
We observe that the dilaton is obviously left unchanged and that
the equation for the electromagnetic field is (formally) the same as in SF.
Indeed, since
$\nabla_i(e^{-a\,\phi}\,F^{ij})
\equiv\partial_i(\sqrt{-g}\,e^{-a\,\phi}\,[F^{ij}]_{EF})
=\partial_i(\sqrt{-G}\,e^{-a\,\phi}\,[F^{ij}]_{SF})$
from (\ref{eq_s}) and (\ref{eq_e}), solutions of Maxwell's equations
in the two frames must be related by
\begin{eqnarray}
e^{2\,\phi}\,\left[F^{ij}\right]_{SF}=\left[F^{ij}\right]_{EF}
\ .
\end{eqnarray}
Therefore the physical (covariant) components of the electromagnetic field
are the same in both frames:
\begin{eqnarray}
\left[F_{ij}\right]_{SF}=\left[F_{ij}\right]_{EF}
\ .
\label{F}
\end{eqnarray}
Furthermore, the uncompactified degrees of freedom of the string do not
move along a geodesic of the metric $g_{ij}$, and the scalar curvatures 
are in general different in the two frames because of the dilaton,
\begin{eqnarray}
&&
R_{(G)}=2\,\left(\nabla\phi\right)^2-3\,\nabla^2\phi
\nonumber \\
&&
R_{(g)}={1\over 2}\,\left(\nabla\phi\right)^2
\ .
\label{R}
\end{eqnarray}
Another general observation is that $S_{EF}$ is invariant under the
following transformation \cite{mirror}
\begin{eqnarray}
{\cal T}\,:\,
\cases{a\to -a \cr
\phi\to-\phi
\ ,}
\label{dual}
\end{eqnarray}
which is not an invariance of $S_{SF}$.
Therefore, although the mapping (\ref{conf}) looks almost trivial at a
first sight, it is clear that the physics in the two frames can be
significantly different.
\par
Which frame is more suitable as a description of the present state of 
our Universe is an open question which will eventually be settled by 
experiment \cite{faraoni}.
The issue raised in the present notes has already been extensively
discussed in the framework of scalar-tensor theories of gravity and 
observable consequences have been deduced mainly in cosmology.
Because of the direct coupling between the dilaton and matter
(in our case the electromagnetic field), both actions in Eqs.~(\ref{s_s})
and (\ref{s_e}) fail to be of the Brans-Dicke type, thus
the equivalence principle does not hold in general.
Specifically, one expects the equivalence principle to be violated
whenever the gradient of the dilaton field is not negligible.
However, such violations might be allowable provided they occurred
far in the past, {\em e.g.} in the early stages of the Universe 
(see \cite{kolb}, \cite{veneziano} and Refs. therein, \cite{cho})
or take place in regions of space which have not been tested at present.
\par
In the following we will analyze some of the physical implications of 
the differences between the two frames for two black hole geometries, 
in which the gradient of the static dilaton field is appreciably strong
only in a relatively small region of space outside the event horizon.
We consider a charged black hole (RND) \cite{rnd,horo} in
section~\ref{s_rnd} and a rotating black hole with small 
charge-to-mass ratio (KND) \cite{knd} in section~\ref{s_knd} 
and suggest possible experiments.
Finally, in section~\ref{light} we compare the propagation of light on
these black hole backgrounds in the two frames.
\section{RND black holes}
\label{s_rnd}
The line element representing a 4-dimensional charged dilatonic
black hole in EF is \cite{rnd,horo}
\begin{eqnarray}
\left.ds^2\right]_{EF}=-e^{2\,\Phi}\,dt^2+e^{2\,\Lambda}\,dr^2
+R^2\,d\Omega_2^2
\ ,
\label{rnd_ef}
\end{eqnarray}
with $d\Omega_2^2=d\theta^2+\sin^2\theta\,d\varphi^2$ and
\begin{eqnarray}
&&e^{2\,\Phi}=e^{-2\,\Lambda}
=\left(1-{r_+\over r}\right)\,
\left(1-{r_-\over r}\right)^{1-a^2\over 1+a^2}
\nonumber \\
&&R^2=r^2\,\left(1-{r_-\over r}\right)^{2\,a^2\over 1+a^2}
\ .
\end{eqnarray}
One encounters an outer horizon at
\begin{equation}
r_+=M+\sqrt{M^2-(1-a^2)\,Q^2}
\ ,
\label{r+}
\end{equation}
while
\begin{equation} 
r_-=(1+a^2)\,{Q^2\over r_+}
\end{equation}
is a real singularity for $a\not=0$.
The corresponding electromagnetic field has only one non-vanishing
component (the static electric field),
\begin{eqnarray}
&&F_{tr}={Q\over r^2}
\ ,
\label{E}
\end{eqnarray}
and the dilaton field is given by
\begin{eqnarray}
e^{-\phi}=
\left(1-{r_-\over r}\right)^{-{2\,a\over 1+a^2}}=1+{\cal O}(r^{-1})
\ .
\label{phi_rnd}
\end{eqnarray}
We observe that for $a=0$ the above expressions reduce to the pure
Reissner-Nordstr\"om (RN) solution (with a constant $\phi_0=0$ dilaton
field)
and that $M$ and $Q$ represent the physical (ADM) mass and charge
of the black hole.
\par
As mentioned in the Introduction, from (\ref{phi_rnd}) one sees that
the dilaton gradient falls off and becomes negligible sufficiently
far away from $r_-$.
To be more specific, since $e^{\phi}\sim G_N$, one can write the
total force acting on a test mass $m$ as the sum of the force $F_\phi$ 
due to the spatial dependence of $G_N$ and the Newtonian contribution 
$F_N$,
\begin{equation}
F_{tot}\sim-\partial_r\left(G_N\,{M\,m\over r}\right)
=F_\phi+G_N\,{M\,m\over r^2}
\ ,
\end{equation}
and finds that $F_\phi$ becomes of the same order as $F_N$ at
\begin{equation}
r\sim {1+3\,a^2\over 1+a^2}\,r_-
\ .
\end{equation}
This implies that, if one can perform a measurement with the 
precision of one part over $10^N$, one has to go closer than 
$r_c\sim 10^N\,r_-$ to the black hole centre in order to test 
any violation of the equivalence principle.
Further, $r_c$ must lie outside $r_+$ and this gives 
an estimate for the smallest charge-to-mass ratio that the black
hole must posses in order that any deviation can be tested, 
namely
\begin{equation}
{Q\over M}>10^{-N/2}
\ .
\end{equation}
For a solar mass black hole and $N\sim 10$ this means a charge 
of about $10^{34}$ electron charges or $10^{15}$ C.
On the other hand, for a Planck mass black hole with one electron
charge the ratio $Q/M\sim 0.1$ and one only needs a precision
of $N=2$.
\par
Upon transforming to SF one obtains
\begin{eqnarray}
\left.ds^2\right]_{SF}=-\left(1-{r_+\over r}\right)\,
\left(1-{r_-\over r}\right)^{1+2\,a-a^2\over 1+a^2}\,dt^2
+\left(1-{r_+\over r}\right)^{-1}\,
\left(1-{r_-\over r}\right)^{a^2+2\,a-1\over 1+a^2}\,dr^2
+r^2\,\left(1-{r_-\over r}\right)^{2\,a\,(1+a)\over 1+a^2}\,d\Omega_2^2
\ .
\label{rnd_sf}
\end{eqnarray}
An interesting feature of the latter metric is that $G_{\theta\theta}$
(as well as $G_{\varphi\varphi}$) is regular at $r=r_-$ for $a=0$ (RN),
as expected, and also for $a=-1$ \cite{volkov}.
Further, the physical (ADM) mass of the black hole is shifted according
to
\begin{eqnarray}
\left.M_{phys}\right]_{SF}&=&M+{a\,Q^2\over r_+}
=M\,\left(1+{a\,Q^2\over 2\,M^2}\right)
+{\cal O}\left({Q^4\over M^4}\right)
\ .
\label{m_rnd}
\end{eqnarray}
The latter result is however not particularly interesting from the
experimental point of view unless a way can be found to measure
$M$ and $M_{phys}$ {\em separately}.
\par
As models of astrophysical black holes, both the line elements 
(\ref{rnd_ef}) and (\ref{rnd_sf}) are unsatisfactory.
They describe non-rotating spherical objects, but real black
holes are expected to be spinning fast because of angular momentum
conservation during the collapse of the original star
(for further evidence see {\em e.g.} \cite{fabian}).
If this limitation is ignored, the fact that $M$ also appears in the
expression for $r_+$ (and $r_-$ as well, but the latter singularity
is hidden) allows us to propose
an experimental test consisting of the following steps:
\par\noindent
{\it i)} by comparing the acceleration of a charged test particle to the
acceleration of a neutral particle of equal mass the
electric field $F_{tr}$ can be measured;
\par\noindent
{\it ii)} the value of $M_{phys}$ is obtained directly from the
acceleration of the neutral particle at large distance;
\par\noindent
{\it iii)} the radius $r_+$ can be estimated by inferring the largest
distance from which light can escape or by localizing the inner edge
of the accreting disk.
\par\noindent
$F_{tr}$ is still given by Eq.~(\ref{E}), as can be inferred from the 
general relation (\ref{F}).
Thus step {\it i)} allows the computation of  $Q$ and the insertion of
$Q$ into the definition
of $r_+$ in (\ref{r+}) which, together with the measured value of $r_+$
from {\it iii)}, gives $M$.
If $M$ is equal to $M_{phys}$ from {\it ii)}, then EF is the physical 
frame and one might question the stringy origin of the action $S_{g}$; 
in case they are not equal, SF is the physical frame and (\ref{m_rnd}) 
can be used to estimate $a$ \cite{a}.
\par
As a further consequence of the metric being different in the two
frames, we notice that the expression of the area of the surface of
the outer horizon is given respectively by
\begin{eqnarray}
&&\left.{\cal A}\right]_{EF}=4\,\pi\,r_+^{2\over 1+a^2}\,
\left(r_+-r_-\right)^{2\,a^2\over 1+a^2}
\nonumber \\
&&\left.{\cal A}\right]_{SF}=4\,\pi\,r_+^{2\,(1-a)\over 1+a^2}\,
\left(r_+-r_-\right)^{2\,a\,(1+a)\over 1+a^2}
\ ,
\end{eqnarray}
so that 
\begin{equation}
{\left.{\cal A}\right]_{EF}\over\left.{\cal A}\right]_{SF}}
=\left({r_+-r_-\over r_+}\right)^{2\,a\over 1+a^2}<1
\ ,
\end{equation}
for $a>0$ (the same ratio is $>1$ for $a<0$).
As we shall show, this affects the evaporation of the black hole.
\par
\begin{figure}
\leavevmode
\centerline{\epsfysize=200pt\epsfbox{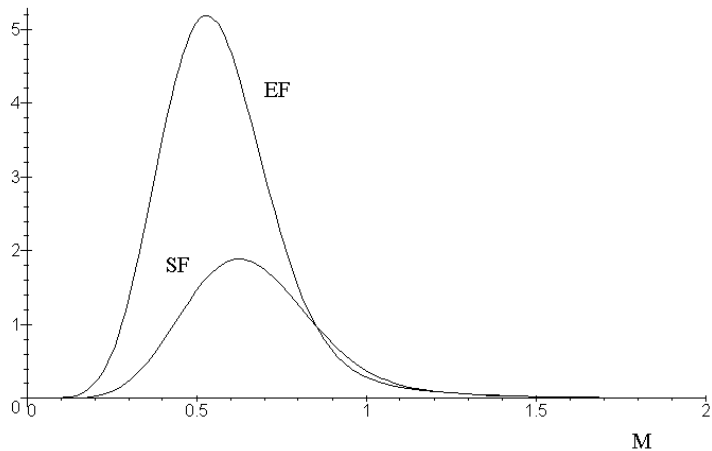}}
\caption{Energy emitted by RND black holes per unit time and ratio $Q/M$ 
fixed in the two frames. 
The mass is in units of the Planck mass.  The vertical scale is arbitrary.}
\label{a}
\end{figure}
\begin{figure}
\leavevmode
\centerline{\epsfysize=200pt\epsfbox{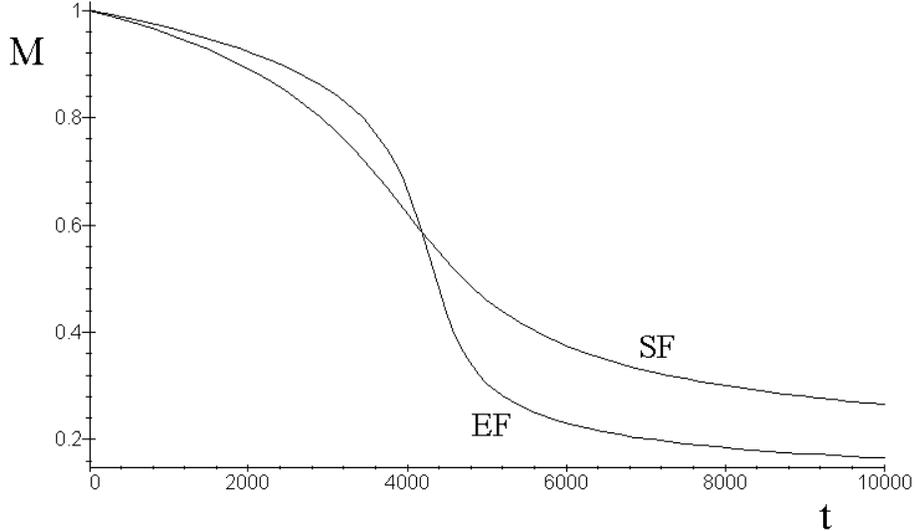}}
\caption{Time evolution of the mass of RND black holes with $M(0)=M_p$
in the two frames.
The time scale is arbitrary.}
\label{b}
\end{figure}

We assume that the area law holds in both frames,
so that the internal degeneracy of the black hole is given
by (see \cite{mg8} and Refs. therein)
\begin{equation}
\Omega\sim e^{{\cal A}/4}
\ ,
\end{equation}
where ${\cal A}$ is the area of the horizon.
Then we study the evolution of the mass of the black hole in time
assuming it can only decrease by emitting Hawking quanta (matter
possibly falling into the black hole and increasing its mass 
will not be considered).
The occupation number density of the Hawking radiation, properly
computed in the microcanonical ensemble where total energy of the
system is conserved and equal to $M_{phys}$, is \cite{mg8}
\begin{equation}
n(\omega)=\sum\limits_{l=1}^{M_{phys}/\omega}
{\Omega(M_{phys}-l\,\omega)\over\Omega(M_{phys})}
\ .
\label{n_o}
\end{equation}
For the sake of simplicity we specialize to the case $a=1$ and also 
assume that the ratio $x\equiv Q/M$ is small and remains constant 
along the evaporation.
Thus we find for the area of the horizon
\begin{eqnarray}
{\cal A}\simeq 4\,\pi\,M^2\,\left(1-f\,x^2\right)
\ ,
\end{eqnarray}
where $M$ is now the physical mass as measured in each frame and
$f=1/2$ in EF ($f=3/2$ in SF).
The occupation number (\ref{n_o}) becomes
\begin{equation}
n_f(\omega)\sim\sum\limits_{l=1}^{M/\omega}
\left[e^{4\,\pi\,l^2\,\omega^2-8\,\pi\,M\,l\,\omega}\right]^{(1-f\,x^2)}
\ .
\label{n_f}
\end{equation}
from which the energy emitted per unit time can be computed according
to
\begin{equation}
{dM\over d\tau}\sim 
-{\cal A}\,\int d\omega\,\omega^3\,\Gamma(\omega)\,n_f(\omega)
\ .
\label{dmdt}
\end{equation}
We have numerically integrated the above expression and the results
for the two frames are shown in Fig.~\ref{a} for $\Gamma=1$ and
$x=1/2$ (a relatively large ratio chosen for the purpose of stressing 
the difference between the two frames).
From that figure one sees that for large values of $M$ the emission
is higher in SF, thus leading to a faster decay, but then the curve
in EF overcomes the curve in SF for values around the Planck mass
$M_p\equiv\sqrt{\hbar\,c/G_N}$ and smaller.
Both curves reach a maximum and then vanish for zero mass, a feature
which is a direct consequence of the use of the microcanonical approach
(that is energy conservation).
The time evolutions which result from Eq.~(\ref{dmdt}) are shown in Fig.~\ref{b}.
When the mass $M$ is of the order of the Planck mass (or bigger), 
Eq.~(\ref{dmdt}) can be approximated by the thermal distribution 
with inverse temperature $\beta=8\,\pi\,M\,(1-f\,x^2)$.
This behaviour would lead to a complete evaporation in a finite time.
However, as can be seen from Fig.~\ref{b}, after the maximum slope is
reached the curves switch to a power law decay and fall to zero in 
infinite time (see Ref.~\cite{mg8} for the details).
The difference between the two frames is that $M$ is smaller in SF
than in EF during the thermal phase but falls off less rapidly in SF than in EF
during the late stages.
\vspace{10pt}
\section{KND black holes}
\label{s_knd}
In Ref.~\cite{knd} a new solution of the Einstein-Maxwell dilatonic
equations (\ref{eq_e}) was found in EF which represents a rotating,
charged black hole with static dilaton field in the small charge-to-mass
($Q/M\ll 1$) approximation.
In a subsequent paper \cite{kndw} we showed that the corresponding
metric can be simplified by shifting the radial coordinate $r$ and
suitably redefining the parameters $M$ (mass), $J$ (angular momentum)
and $Q$ (charge) to make them equal the corresponding physical (ADM)
quantities $M_{phys}$, $J_{phys}$ and $Q_{phys}$ which determine the
Newtonian motion in the asymptotically flat region.
One then has
\begin{equation}
g_{ij}=g_{ij}^{KN}+{\cal O}\left({Q^4\over M^4}\right)
\ ,
\end{equation}
where $g_{ij}^{KN}$ is the usual Kerr-Newman (KN) metric which we write
in Boyer-Lindquist coordinates \cite{chandra},
\begin{eqnarray}
ds^2_{KN} = -\sqrt{\Delta}\,\sin\theta\,\left[
\chi\,d\varphi^2-{1\over\chi}\,(dt-\omega\,d\varphi)^2\right]
+\rho^2\,\left[{(dr)^2\over\Delta}+(d\theta)^2\right]
\ ,
\label{g_ij}
\end{eqnarray}
in which ($\alpha\equiv J/M$)
\begin{eqnarray}
\chi&=&{\sqrt{\Delta}\,\sin\theta\over\Psi}
\nonumber \\
\Delta&=&r^2-2\,M\,r+\alpha^2+Q^2
\nonumber \\
\rho^2&=&r^2+\alpha^2\,\cos^2\theta
\nonumber \\
\Psi&=&-{\Delta-\alpha^2\,\sin^2\theta\over\rho^2}
\nonumber \\
\omega&=&-\alpha\,\sin^2\theta\,[1+\Psi^{-1}]
\ .
\label{KN-metric}
\end{eqnarray}
This also implies that the causal structure is not affected by
the presence of the dilaton at that order in EF.
In fact one still has two horizons for $\Delta=0$, that is
\begin{equation}
r_\pm=M \pm\sqrt{M^2-\alpha^2- Q^2}
\ .
\end{equation}
\par
However the presence of a non-zero dilaton field,
\begin{eqnarray}
\phi=-a\,{r\over\rho^2}\,{Q^2\over M}
\ ,
\label{p0}
\end{eqnarray}
affects the electric and magnetic field potentials $A$ and $B$
(see Ref.~\cite{knd} for the definitions),
\begin{eqnarray}
A&=&Q\,{r\over\rho^2}\,\left[
1-\left({1\over2\, r}+{r\over\rho^2}\right)\,
{a^2\,Q^2\over 3\,M}\right]
\nonumber \\
B&=&-Q\,\alpha\,{\cos\theta\over\rho^2}\,
\left[1-\left({1\over 2\,M}-{r\over\rho^2}\right)\,
{a^2\,Q^2\over3\,M}\right]
\ ,
\end{eqnarray}
where the terms proportional to $Q^2$ inside
the brackets correspond to the corrections with respect to
the KN potentials \cite{chandra}.
These corrections also affect the electric and magnetic fields
in the limit $r\to\infty$,
\begin{eqnarray}
{\cal E}_{\hat{r}} &\approx& {Q\over{r^2}}
\nonumber \\
{\cal E}_{\hat{\theta}} &\approx&
-{2\,\alpha^2\, Q\over{r^4}}\,\sin\theta\,\cos\theta
\nonumber \\
{\cal B}_{\hat{r}} &\approx&
{2\,\alpha\, Q\over{r^3}}\,\cos\theta
\,\left[1 - {a^2\,Q^2\over{6\,M^2}}\right]
\nonumber \\
{\cal B}_{\hat{\theta}} &\approx&
{\alpha\, Q\over{r^3}}\,
\sin\theta\left[1 - {a^2\,Q^2\over{6\,M^2}}\right]
\equiv{\mu_{phys}\over r^3}\,\sin\theta
\ ,
\label{E_knd}
\end{eqnarray}
where ${\cal E}_{\hat a}$ and ${\cal B}_{\hat a}$ ($\hat a
=\hat r, \hat\theta$) are respectively
the electric and the magnetic field components with respect to the
usual spherical coordinate tetrad basis \cite{misner}.
Eq.(\ref{E_knd}) shows a relative shift in the intensity of
the field ${\cal B}$ with respect to ${\cal E}$.
This is also the source for the anomalous gyromagnetic ratio
\begin{eqnarray}
[g]_{EF}=2\,{\mu_{phys}\,M_{phys}\over Q_{phys}\,J_{phys}}\simeq
2\,\left[1-{a^2\,Q^2\over 6\, M^2}\right]
\ ,
\label{gyro_ef}
\end{eqnarray}
which for the KND black hole cannot be greater than 2 and is equal
to 2 for the KN black hole \cite{strau}.
\par
An obvious consequence of the above form of the solution is that
the geodesic motion of neutral particles, which do not couple directly
to the dilaton, are unaffected by the presence of static dilaton field
(up to order $Q^3/M^3$).
As we have mentioned in the Introduction, this should not  be the case
for fundamental strings, from whose action $S_{SF}$ has been derived by
compactifying extra dimensions.
To see that this is indeed the case we compute the SF metric
corresponding to $g_{ij}^{KN}$ with the dilaton field given in
(\ref{p0}), thus obtaining
\begin{eqnarray}
G_{ij}=g_{ij}^{KN}\,\left(1-a\,{r\over\rho^2}\,{Q^2\over M}\right)
+{\cal O}\left({Q^4\over M^4}\right)
\ .
\label{G}
\end{eqnarray}
Then we observe that (\ref{G}) cannot be remapped into
$g_{ij}^{KN}$ by a change of coordinates since the curvature scalars
are different in the two frames.
In fact, from (\ref{R}) one obtains
\begin{eqnarray}
&&
R_{(G)}\approx 12\,a\,{Q^2\over M}\,\alpha^2\,\cos 2\theta\,{1\over r^3}
\nonumber \\
&&
R_{(g)}\approx {a^2\over 2}\,{Q^4\over M^2}\,{1\over r^4}
\ ,
\label{R1}
\end{eqnarray}
where, for the sake of simplicity, we have displayed only the behaviors
for large $r$.
\par
Now we analyze further physical differences between the two frames.
To start with, since
\begin{equation}
G_{tt}\approx-\left[1-{2\,M\over r}\,\left(1+{a\,Q^2\over 2\,M^2}\right)
\right]
\ ,
\end{equation}
for $r\to+\infty$, the ADM mass in SF is shifted to
\begin{equation}
\left[M_{phys}\right]_{SF}=\left[M_{phys}\right]_{EF}\,
\left(1+{a\,Q^2\over 2\,M^2}\right)
\ ,
\end{equation}
which is the same expression that was obtained on expanding
the physical mass in $Q/M$ for RND.
Thus, the same experimental test described in Section~\ref{s_rnd} 
after Eq.~(\ref{m_rnd}) can be repeated for the present case.
We remind the reader here that the KND metric is
a more realistic candidate for the description of astrophysical black holes,
since it describes black holes which possess angular momentum.
\par
We notice that
\begin{eqnarray}
G_{t\varphi}\approx-2\,{\alpha\,M\over r}\,\sin^2\theta
\equiv-2\,{J_{phys}\over r}\,\sin^2\theta
\ ,
\end{eqnarray}
therefore $J_{phys}$ is left unchanged by the conformal mapping as well
as $\mu_{phys}$, see (\ref{F}) and (\ref{E_knd}).
The gyromagnetic ratio in SF is then given by
\begin{eqnarray}
[g]_{SF}\simeq 2\,\left[1+{a\,Q^2\over 2\,M^2}\,\left(1-{a\over 3}\right)
\right]
\ .
\label{gyro_sf}
\end{eqnarray}
The above expression for $[g]_{SF}$ shows a remarkable difference
with respect to
that of EF, that is $[g]_{SF}$ can be either smaller than $2$ (for $a<0$
or $a>3$),
equal to $2$ for ($a=0,3$) or greater than $2$ (for $0<a<3$).
The existence of the latter case provides another way of testing which
frame is the physical one.
In fact, since $[g]_{EF}$ can be at most equal to $2$, the measurement
of a value greater than $2$ for the gyromagnetic ratio of a black hole
would prove that Physics has to be described in SF (this should indeed
be the case \cite{a}).
On the other hand, the measurement of any value smaller than $2$,
although crucial for proving the existence of static dilaton field,
would not suffice for discriminating between EF and SF, unless an
independent way of measuring $a$ along with the mass and charge of the
black hole can be found.
\section{Light propagation}
\label{light}
In this section we study the differences which emerge in the propagation
of electromagnetic signals in the two frames.
\par
We start from the approximation of geometric optics and look at
the deflection angle of a null ray impinging upon the black hole from
far away and then escaping to $r=+\infty$.
Since the relation between SF and EF is given by a conformal
transformation of the metric, the naive expectation is that no changes
occur.
In fact it is easy to prove that the picture is exactly the same in both
frames, and the eikonal path followed by any null ray is 
unaffected by the choice of the frame.
\par
We consider for the metric in EF a generic form $g_{ij}=g_{ij}(r,\theta)$,
with only one possible off-diagonal term ($g_{t\varphi}$), which can
be easily specialized to both RND and KND.
The Lagrangian for a null particle moving on such a metric and in the
equatorial plane $\theta=\phi/2$ is given by \cite{chandra}
\begin{eqnarray}
2\,{\cal L}=g_{tt}\,\dot t^2+2\,g_{t\varphi}\,\dot t\,\dot\varphi
+g_{\varphi\varphi}\,\dot\varphi^2+g_{rr}\,\dot r^2=0
\ ,
\end{eqnarray}
where a dot denotes the derivative with respect to an affine parameter
$\lambda$.
The conserved momenta are
\begin{eqnarray}
&&p_t=g_{tt}\,\dot t+g_{t\,\varphi}\,\dot\varphi\equiv E
\nonumber \\
&&p_\varphi=g_{tt}\,\dot\varphi+g_{t\,\varphi}\,\dot t\equiv L
\ ,
\end{eqnarray}
corresponding respectively to the energy and angular momentum of the null
particle.
In particular, from the conservation of $p_\varphi$ one obtains
\begin{eqnarray}
\left[\dot\varphi\right]_{EF}=D^{-1}\left(g_{tt}\,L-g_{t\varphi}\,E\right)
\ ,
\end{eqnarray}
with $D\equiv g_{tt}\,g_{\varphi\varphi}-g_{t\varphi}^2$,
and, after substituting for $\dot t$ and $\dot\varphi$ in the
Lagrangian,
\begin{eqnarray}
\left[\dot r^2\right]_{EF}={D^{-1}\over g_{rr}}\,
\left(g_{tt}\,L^2+g_{\varphi\varphi}\,E^2-2\,g_{t\varphi}\,E\,L\right)
\ .
\end{eqnarray}
From the above expressions for $\dot\varphi$ and $\dot r$ one finally
obtains
\begin{eqnarray}
\left[{d\varphi\over dr}\right]_{EF}=
\left[{d\varphi\over d\lambda}\right]_{EF}\,
\left[{d\lambda\over dr}\right]_{EF}
=\sqrt{g_{rr}\over D}\,
{g_{tt}\,L-g_{t\varphi}\,E\over
\sqrt{g_{tt}\,L^2+g_{\varphi\varphi}\,E^2-2\,g_{t\varphi}\,E\,L}}
\ .
\label{df/dr}
\end{eqnarray}
The deflection angle is then given by
\begin{eqnarray}
\Delta\varphi=
2\,\left(\varphi(\bar r)-\varphi(\infty)\right)-\pi
\ ,
\end{eqnarray}
where
\begin{eqnarray}
\varphi(r)=\int^r \left[{d\varphi\over dr}\right]_{EF}\,dr
\ ,
\end{eqnarray}
and $\bar r$ is the minimum radial coordinate reached by the particle,
that is
\begin{eqnarray}
\left.\left[\dot r\right]_{EF}\right|_{\bar r}=0
\ .
\label{turn}
\end{eqnarray}
Explicit expressions can then be obtained for the two kinds of black holes
by simply substituting in the corresponding metric elements \cite{kndw2}.
\par
Switching to SF, one obtains
\begin{eqnarray}
&&\left[\dot r\right]_{SF}=e^{-\phi}\,\left[\dot r\right]_{EF}
\nonumber \\
&&\left[\dot\varphi\right]_{SF}=e^{-\phi}\,\left[\dot\varphi\right]_{EF}
\ .
\end{eqnarray}
Thus, since $e^{-\phi}$ is a positive function, one finds that the
turning point at which the particle stops approaching the black hole 
has radial coordinate $\bar r$ which does not depend on the frame.
Further
\begin{eqnarray}
\left[{d\varphi\over dr}\right]_{SF}=
\left[{d\varphi\over dr}\right]_{EF}
\ .
\end{eqnarray}
One can then conclude that $\Delta\varphi$ is frame-independent and that
eikonal trajectories of null waves are exactly the same in both EF and SF.
In particular, for KND this means that to lowest order in $(Q/M)^2$
eikonals
do not sense the dilaton at all \cite{kndw2}.
\par
Of course the above conclusion does not preclude the possibility of
detecting other differences in the propagation of light waves in SF 
with respect to EF.
To begin with, we notice that, although the radial coordinate of
the turning points defined by Eq.~(\ref{turn}) is frame-independent,
the actual proper distance to an observer placed at $r=r_o$ depends 
on the frame.
If we consider the RND case and $a=1$, we see that the difference 
between proper distances to $r_o$ is given by
\begin{equation}
\left.l\right]_{EF}=\left.l\right]_{SF}
+r_-\,\ln\left({r_o-r_+\over \bar r-r_+}\right)
\ ,
\end{equation}
which would result in a delay for the time of flight from $\bar r$  
to $r_o$ in EF with respect to SF.
\par
\begin{figure}
\leavevmode
\centerline{\epsfysize=60pt\epsfbox{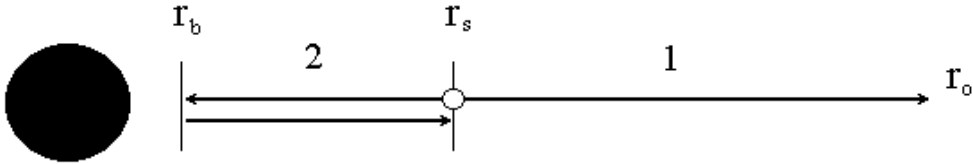}}
\caption{Simple model of {\em reverberation}.
Explanation is given in the text.}
\label{c}
\end{figure}

A possible way of detecting such a time delay is displayed in Fig.~\ref{c}
where one considers a source of light at $r=r_s$ which emits both
towards the observer placed at $r_o$ (ray 1) and towards the black hole
(ray 2).                                              
The latter ray then bounces back at $r_b$ and reaches the observer
with a delay with respect to ray 1 given by twice the time it takes to
go from the source to $r_b$ (this is a simple model for the so called
{\em reverberation} in the accreting disk, see \cite{fabian2}).
In EF this delay is given by (again assuming RND with $a=1$)
\begin{eqnarray}
\left.\tau\right]_{EF}\sim 2\,\left[r_s-r_b
+r_+\,\ln\left({r_s-r_+\over r_b-r_+}\right)\right]
\ ,
\end{eqnarray}
while in SF one has
\begin{eqnarray}
\left.\tau\right]_{SF}\sim\left.\tau\right]_{EF}  
-2\,r_-\,\ln\left({r_s-r_+\over r_b-r_+}\right)
\ .
\end{eqnarray}
\par
The difference between the metrics in the two frames also affects
the red-shift $z$ of waves emitted at $r_s$.
For instance, in RND with $a=1$ one has
\begin{eqnarray}
&&\left.z\right]_{EF}=-{r_+\over r_s}=-{2\,M\over r_s}
\nonumber  \\
&&\left.z\right]_{SF}=\left.z\right]_{EF}-{r_-\over r_s}+
{r_-\,r_+\over r_s^2}=\left.z\right]_{EF}
-{Q^2\over r_s}\,\left({1\over M}-{2\over r_s}\right)
\ .
\end{eqnarray}
\par
Finally, there are effects that cannot be accounted for without
fully considering the wave nature of electromagnetic radiation.
In fact, the electromagnetic waves couple directly to the dilaton
(as can be seen in the action, see also Refs.~\cite{kndw,kndw2} for
explicit expressions in EF) and, although at leading orders the eikonal
trajectories look the same in both frames, the intensity of the waves is
different because of the different metric backgrounds.
From (\ref{F}) one can estimate the intensity of electromagnetic
radiation produced in scattering events involving other fields in the 
model according to $I\sim |\vec E|^2+|\vec B|^2\sim F^2$.  
Therefore from (\ref{F}) one gets
\begin{eqnarray}
[I]_{SF}\sim e^{-2\,\phi}\,\left[F^2\right]_{EF}\sim e^{-2\,\phi}\,
[I]_{EF}
\ .
\end{eqnarray}
This implies $[I]_{SF}\sim [I]_{EF}\,\left(1+{\cal O}(r^{-1})\right)$ 
and the difference is not appreciable when the scatterings occur far 
away from the hole.
\section{Conclusion}
The resolution of the issue of which frame is more suitable for the
description of physical processes will have to wait at least until 
the next generation of astrophysical instruments.  
The measurement of the gyromagnetic ratio of a black hole will be a 
sensitive test of SF vs. EF, but precise measurements of the 
magnetic moment, the mass, the charge and the angular momentum of 
the black hole are necessary for the determination of this ratio.
Existing telescopes are incapable of making such measurements, but 
future instruments will no doubt be able to `see' close enough to 
the horizon of the black hole to determine these quantities with 
sufficient precision.
Measurement of the intensity of electromagnetic radiation passing 
near the horizon of a black hole would also, in principle, be a very 
sensitive way of distinguishing between the two frames.
This would require measuring the intensity of light from the stellar
component of a black hole binary system for two different locations 
of the star -- once when the star is occulting and once when it is
eclipsing the black hole.  
The mass and charge of the black hole would also have to be known, as
well as the angular momentum if frame dragging effects are to be taken
into account.  
Although such binary systems are known to exist, precise measurement 
of the appropriate parameters is beyond present capabilities and will
have to wait until the next generation of instruments become operative.
\acknowledgments
R. C. thanks M. Giovannini for useful discussions.  
This work was supported in part by the U.S. Department of Energy under 
Grant No. DE-FG02-96ER40967.
%

%

\begin{thebibliography}{99}
%
\bibitem{gsw}
M.B. Green, J.H. Schwarz, E. Witten, {\em Superstring theory},
Cambridge University Press, Cambridge (1987).
%
\bibitem{lovelace}
C. Lovelace, {\em Phys. Lett.} {\bf B135} (1984) 75.
%
\bibitem{callan}
E.S. Fradkin and A.A. Tseytlin, {\em Nucl. Phys.} {\bf B261} (1985) 1;
C.G. Callan, D. Friedan, E.J. Martinec and M.J. Perry,
{\em Nucl. Phys.} {\bf B262} (1985) 593;
A. Sen, {\em Phys. Rev.} D {\bf 32} (1985) 2102.
%
\bibitem{kolb}
E.W. Kolb and M.S. Turner, {\em The Early Universe}, Addison-Wesley,
Redwood City, 1990.
%
\bibitem{horo}
G.T. Horowitz and A. Strominger, {\em Nucl. Phys.} {\bf B360} (1991) 197.
%
\bibitem{a}
From string theoretical computations one has $a=1$.
However, this value might change and the dilaton might acquire 
a mass from quantum corrections.
%
\bibitem{mirror}
If the fields contain only even powers of $a$, then one expects that 
there exists a transformation which maps solutions with $\phi$ 
to solutions with $-\phi$.
For instance, the solution found in Ref.~\cite{knd} in the weak 
dilaton regime is connected to a solution in the strong dilaton 
regime by a generalization of the symmetry of Ernst's equations
\cite{chandra} which contains the second transformation in (\ref{dual}).
%
\bibitem{faraoni}
For a review of the issues involved in determining which conformal frame
is the physical one see V. Faraoni, E. Gunzig and P. Nardone, 
{\it Conformal transformations in classical gravitational theories and
in cosmology}, gr-qc/9811047
%
\bibitem{veneziano}
G. Veneziano, {\em A simple/short Introduction to Pre-Big-Bang
Physics/Cosmology}, preprint CERN-TH/98-43, hep-th/9802057.
%
\bibitem{cho}
Y.M. Cho and Y.Y. Keum, Class. Quantum Grav. {\bf 15} (1998) 907.
%
\bibitem{rnd}
G.W. Gibbons and K. Maeda, {\em Nucl. Phys.} {\bf B298},
741 (1988).
%
\bibitem{knd}
R. Casadio, B. Harms, Y. Leblanc and P.H. Cox, Phys. Rev. D
{\bf 55}, 814 (1997).
%
\bibitem{volkov}
For static spherically symmetric solutions of Einstein-Yang-Mills-Dilaton 
equations with $a\not=0$ it has been proven that there cannot exist an 
inner horizon in EF, see O. Sarbach, N. Straumann and M. S. Volkov,
{\em Internal structure of Einstein-Yang-Mills-Dilaton black holes},
preprint gr-qc/9709081.
%
\bibitem{fabian}
A. C. Fabian, Astron. Geoph. {\bf 38}, 10 (1997). 
%
\bibitem{mg8}
R. Casadio and B. Harms, {\em Microfield Dynamics of Black Holes},
to appear in Phys. Rev. D, preprint gr-qc/9712017.
%
\bibitem{kndw}
R. Casadio, B. Harms, Y. Leblanc and P.H. Cox, Phys. Rev. D 
{\bf 56}, 4948 (1997).
%
\bibitem{chandra}
S. Chandrasekhar,
{\it The Mathematical Theory of Black Holes},
Oxford University Press, Oxford (1983).
%
\bibitem{misner}
C.W. Misner, K.S Thorne and J.A. Wheeler,
{\it Gravitation}, W.H. Freeman and Co., San Francisco,
1973.
%
\bibitem{strau}
N. Straumann,
{\it General Relativity and Relativistic Astrophysics},
Springer-Verlag, Berlin (1984).
%
\bibitem{kndw2}
R. Casadio and B. Harms,
{\em Perturbations in the Kerr-Newman Dilatonic Black Hole Background: 
Maxwell Waves, the Dilaton Background and Gravitational Lensing},
to appear in Phys. Rev. D, preprint gr-qc/9804045.
%
\bibitem{fabian2}
C. S. Reynolds, A. J. Young, M. C. Begelman and A. C. Fabian,
{\em X-ray iron line reverberation from black hole accretion disk},
preprint and private communication.
%
\end{thebibliography}
\end{document}